\documentstyle[11pt]{article}
\def\pl#1#2#3{Phys.~Lett.~{\bf {#1}B} (19{#2}) #3}
\def\np#1#2#3{Nucl.~Phys.~{\bf B{#1}} (19{#2}) #3}

\def\pr#1#2#3{Phys.~Rev.~{\bf D{#1}} (19{#2}) #3}

\def\cmp#1#2#3{Commun.~Math.~Phys.~{\bf {#1}} (19{#2}) #3}
\def\jmp#1#2#3{J.~Math.~Phys.~{\bf {#1}} (19{#2}) #3}

\newcommand{\eq}{\begin{equation}}
\newcommand{\en}{\end{equation}}
\newcommand{\eqn}{\begin{eqnarray}}
\newcommand{\enn}{\end{eqnarray}}

\newcommand{\beq}{\begin{equation}}
\newcommand{\eeq}{\end{equation}}
\def\esev{$E_{7(7)}$}
\def\esix{$E_{6(6)}$}
\def\eja{$J_3^{{\bf O}}$}

\def\j2{$J_2^{{\bf C}}$}
\def\f4{$F_{4(4)}$}
\begin{document}
\begin{titlepage}
\begin{flushright}
  CERN-TH-97-187 \\
  PSU-TH-188\\
\end{flushright}
\begin{center}
{\bf Orbits of Exceptional Groups, Duality and BPS States in 
String Theory} \\
\vspace{1cm} 
{\bf Sergio Ferrara\footnote{Work supported in part by EEC under
TMR  contract ERBFMRX-CT96-00 (LNF Frascati, INFN, Italy) and DOE grant
DE-FGO3-91ER40662, DE-FGO2-96ER40559 \newline
e-mail: ferraras@vxcern.cern.ch}
 and Murat G\"{u}naydin} \footnote{Work supported in part 
by the National Science Foundation under Grant Number PHY-9631332.\newline
Permanent address: Penn State University, Physics Dept. University Park,
PA 16802 
\newline e-mail: murat@phys.psu.edu}  \\
Theory Division\\
CERN \\
CH-1211 Geneva 23\\
\vspace{1cm}
{\bf Abstract}
\end{center}

We give an invariant classification of orbits of the fundamental 
representations of exceptional groups $E_{7(7)}$ and $E_{6(6)}$
which classify BPS states in string and M theories toroidally compactified
to $d=4$ and $5$. The exceptional Jordan algebra and the exceptional 
Freudenthal triple system and their cubic and quartic invariants play
a major role in this classification. 
 The cubic and quartic invariants
correspond to the black hole entropy in $d=5$ and $d=4$, respectively.
The classification of BPS states preserving different numbers of supersymmetries 
is in close parallel to the classification of the little groups and 
the orbits of timelike, lightlike and  space-like vectors in Minkowski space. 
The orbits of  BPS black holes in $N=2$ Maxwell-Einstein supergravity
theories in $d=4$ and $d=5$ with symmetric space geometries are also
classified including the exceptional $N=2$ theory that has $E_{7(-25)}$ and
$E_{6(-26)}$ as its symmety in the respective dimensions.
\end{titlepage}

\renewcommand{\theequation}{\arabic{section} - \arabic{equation}}
\section{Introduction}
\setcounter{equation}{0}
        The exceptional groups \esev \/ and \esix \/ appear as
 duality symmetries
\cite{cj,ec} of 
the low energy actions and their discrete subgroups as symmetries of the
non-perturbative BPS spectrum of string  and M theories in $d=4$ and $5$  
preserving $N=8$ supersymmetry \cite{ht}. The charges of the extremal 
BPS black holes can be assigned to the fundamental representations 
of the exceptional groups \esev \/ and \esix \/ which are 56 and 27
 dimensional
,respectively. The entropy of these black holes in $d=5$ and $d=4$ 
is given by the square root of the cubic and quartic invariants of 
\esix \/ and \esev \/ , respectively \cite{fk,kk}. However, the charge
 configurations 
must satisfy additional restrictions depending on the number of
 supersymmetries 
preserved. In fact, the eigenvalues of the
central charge matrix must be degenerate when more than one supersymmetry
is preserved by the black hole solution. 
 These constraints were recently investigated in terms of
 a certain set 
of invariant conditions on the representation \cite{fm}. In this paper 
we give a classification of such BPS states in terms of orbits of 
\esix \/ and \esev \/ in the corresponding representation. We then extend
our results to $N=2$ Maxwell-Einstein supergravity theories (MESGT) 
in $d=5$ and
$d=4$ dimensions  whose
scalar manifolds are symmetric spaces and determine 
the orbits of their BPS black holes.  
In particular we give the orbits of the exceptional $N=2$ MESGT 
that has $E_{6(-26)}$ and $E_{7(-25)}$
as its symmetry in five and four dimensions, respectively.

\section{Jordan Algebras , Exceptional Groups and Their Orbits}
\setcounter{equation}{0}

The cubic invariant $I_3$ in the 27 dimensional representation of $E_6$
can be identified with the cubic norm of the  exceptional Jordan
algebra $J_3^{{\bf O}}$ of $3\times 3 $ hermitian matrices
over the composition algebra of octonions ${\bf O}$ with the
symmetric Jordan product 
\eq
j_1 \circ j_2 = j_2 \circ j_1
\en
that satisfies the Jordan identity \cite{jvw,nj,mg75,gpr,gstpl,gstnp}
\eq
j_1 \circ ( j_2 \circ j_1^2 ) = (j_1 \circ j_2 ) \circ j_1^2 
\en 
A generic element $j$
of \eja \/ has the form
\begin{equation}
j=  \left( \begin{array}{ccc} 
\alpha_1 & o_3 & o_2^* \\
 o_3^* & \alpha_2 & o_1 \\
o_2 & o_1^* & \alpha_3 
\end{array}
\right)
\end{equation}
where $\alpha_i$ take values over the underlying field which
we take to be real numbers ${\bf R}$ and $o_i$ ($ i=1,2,3$) are elements
of ${\bf O}$. The norm of an octonion $o$ is defined as
\eq
N(o)= o o^* = o^* o
\en
where $*$ denotes octonion conjugation. There are different forms of 
the composition algebra of
 octonions. For the division algebra of real octonions the norm
is invariant under $O(8)$ and for split octonions the norm is invariant
under $O(4,4)$. For $N=8$ supergravity the relevant form of \eja \/ is the
one defined over the split octonions  and for the exceptional $N=2$ 
Maxwell-Einstein supergravity \cite{gstpl} it is the one defined over real
 octonions.
We shall refer to the algebra defined over split octonions
 as the split exceptional Jordan algebra.
\footnote{We should note that the split exceptional Jordan algebra
 and its associated symmetries first
 appeared in physics literature in attempts to find octonionic realizations
of space-time supersymmetry \cite{mg75}.} 
The automorphism group of the split exceptional Jordan algebra
 is the noncompact $F_{4(4)}$ with maximal
compact subgroup $USp(6) \times USp(2)$ \cite{mg75}. Note that $F_{4(4)}$
is also the isometry group of the quaternionic manifold of a maximal $N=2$ 
matter-Einstein supergravity one can obtain by truncation of the $N=8$
supergravity in $d=5$ \cite{gstpl,adf}.
 The cubic norm $I_3$ of \eja \/ is given by
\eq
 I_3 =\alpha_1 \alpha_2 \alpha_3 -\alpha_1( o_1o_1^*) -\alpha_2( o_2 o_2^*)
- \alpha_3( o_3 o_3^*) +2 Re (o_1 o_2 o_3)
\en
where $Re$ represents the real part of an octonion and satisfies
\eq
Re(o_1 o_2) o_3 = Re o_1 ( o_2 o_3)
\en
The invariance group of the norm form of a Jordan algebra $J$ is referred to
as the reduced structure group \cite{nj} and denoted as $St_0 (J)$.
For the split exceptional Jordan algebra
 the reduced structure group is the exceptional group
$E_{6(6)}$ with a maximal compact subgroup $USp(8)$. We should  also note that
$USp(8)$ is the automorphism group of the $N=8$ Poincare supersymmetry 
algebra in $d=5$ \cite{ec}. An element of \eja \/ can be brought to a
 diagonal
form by an \f4 rotation \cite{gpr} and if we denote the eigenvalues of a generic element
$j$ as $ \lambda_i$  ($ i=1,2,3$) the cubic norm is simply
\eq
I_3(j)= \lambda_1 \lambda_2 \lambda_3
\en

To make the analysis that follows clearer from a physics perspective we shall
make an analogy with Minkowski space $M_4$ and its symmetries following 
\cite{mg75,mg80,mgm}. A four vector in $M_4$ can be represented by
$2\times 2$ matrices $x=x_{\mu}\sigma^{\mu}$ where $\sigma^0=1_2$ and
 $\sigma^i$ $( i=1,2,3)$ 
are the Pauli matrices. As $2\times 2 $ matrices the coordinates $x$ can be considered as
elements of the Jordan algebra $J_2^{{\bf C}}$
of Hermitian matrices over the complex numbers ${\bf C}$ with the symmetric Jordan product
which preserves hermiticity. The automorphism group of \j2 is the covering group
$SU(2)$ of the rotation group which is the analog of $F_{4(4)}$ for \eja \/ . The norm form
of \j2 is quadratic and is given by the ordinary determinant.  The invariance group 
the quadratic norm  of \j2 is the covering group $Sl(2,{\bf C})$ of the Lorentz 
group $SO(3,1)$ which is the analog of \esix \/ for \eja \/. In Minkowski space a vector
is characterized by its norm and the parameters of the corresponding orbits. 
Time-like, space-like and light-like vectors corresponding to positive, negative and 
vanishing norms  have orbits
$\frac{Sl(2,{\bf C})}{SU(2)}$, $\frac{Sl(2, {\bf C})}{SU(1,1)}$ and
 $\frac{Sl(2, {\bf C})}{E_2}$ , respectively. Similarly, we can characterize
the elements of \eja \/ by their norms and the parameters of their orbits.
The generic orbit corresponding to a non-vanishing norm $I_3(j)$ has the
26 dimensional orbit
\eq
\frac{E_{6(6)}}{F_{4(4)}}
\en
In contrast to the Minkowskian case the little groups of 
"space-like" and  "time-like" vectors are the same in the case of \eja \/ since 
its norm is cubic. \footnote{As we shall see later the orbits of "time-like" and
"space-like"  vectors are quite different in  four dimensions where  the  invariant 
norm form is quartic!} As for "light-like" elements $j$ of \eja \/
with $I_3(j)=0$ there exist two distinct orbits depending on whether one or two
of the eigenvalues of $j$ vanish. The generic light-like orbit 
corresponding to a single vanishing eigenvalue is given by the 26
dimensional coset space
\eq
\frac{E_{6(6)}}{O(5,4)\odot T_{16}}
\en
where $\odot$ stands for semidirect product and $T_{16}$ 
are translations corresponding to the spinor
 representation of $O(5,4)$ . The $T_{16}$  decomposes as
$T_{8_s}\oplus T_{8_c}$ under the $O(4,4)$ subgroup where
$8_c$ and $8_s$ are the two spinor representations of $O(4,4)$.

The critical light-like orbit corresponds to an element $j$
with two vanishing eigenvalues and is given by the 17 dimensional
space 
\eq
\frac{E_{6(6)}}{O(5,5)\odot T_{16}}
\en
where $T_{16}$ is an Abelian subgroup  corresponding to the 
spinor representation of $O(5,5)$.
We should note that the distinction between generic and critical light-like
orbits does not exist in the Minkowskian case since the norm is
quadratic in that case.\footnote{ In $d=4$ where the cubic norm is 
replaced by a quartic norm an even richer structure exists as we shall
see later.}

As is well-known the invariance group of the light-cone in Minkowski
space $M_4$ is the conformal group $SO(4,2)$ which acts non-linearly.
In fact the Minkowski space $M_4$ is simply the quotient space
\eq
\frac{SO(4,2)}{[SO(3,1)\times O(1,1)] \odot T_4}
\en
When we represent the coordinates of $M_4$ in terms of hermitian
$2\times 2 $ matrices the action of the conformal group can be represented 
as a linear fractional group. As such the conformal group can be interpreted
as the linear fractional group of quaternions \cite{feza}.
However, if we think of the $2\times 2$ hermitian matrices as elements of the
Jordan algebra $J_2^{\bf C}$ the conformal group becomes the linear
fractional group of  Jordan algebra \j2 \cite{mk,mg75,mg80} which generalizes to
 all Jordan algebras and Jordan superalgebras \cite{mk,mg80,mgm}.
The invariance group of the light-cone of \eja \/ defined by the condition
$I_3(j)=0$ is the noncompact exceptional group \esev \/ which acts as the linear
fractional group of \eja \cite{mk,mg75,mgm}. This implies that the 27
dimensional space of \eja \/ can be regarded as the quotient space
\eq
\frac{E_{7(7)}}{[E_{6(6)}\times O(1,1)]\odot T_{27}}  
\en

The above examples of linear fractional group actions are particular cases
of the general nonlinear  actions of noncompact groups $G$ whose Lie algebras
$g$ admit a three grading with respect to a maximal rank subalgebra $g^0$
\eq
g = g^{-1} \oplus g^0 \oplus g^{+1}
\en
In such cases there exists a nonlinear action of $G$ on the grade 
$+1$ space $g^{+1}$ via fractional linear transformations
\cite{mk,bg}. In the case
of \esev \/, $g^0$ is simply the Lie algebra of \esix $
 \times O(1,1)$ and
$g^{+1}$ is the 27 dimensional subspace corresponding to \eja\/ .
We will comment  on the relevance of the ``conformal'' extensions of
duality groups later.

We now consider the symmetries of superstring or M theories toroidally
compactified to four dimensions with $N=8$ supersymmetry. In this case the
duality group is \esev \/ with maximal compact subgroup $SU(8)$. The compact
subgroup $SU(8)$ acts as the automorphism group of the $N=8$ supersymmetry
algebra. The generic charged vector for a BPS state has 56 components with
a quartic norm $I_4$.   
The 56 dimensional representation space of \esev \/ can be represented as
 elements of the exceptional Freudenthal triple system \cite{hf} which
can be realized as $2\times 2 $ ``matrices''of the form \cite{fts}:

\begin{equation}
q =  \left( \begin{array}{cc} 
\alpha  & x \\
 y & \beta  
\end{array}
\right)
\end{equation}

where $\alpha , \beta \in {\bf R} $ and $x, y$ are elements of \eja \/.
One can define a symmetric four-linear form over the exceptional Freudenthal
system which induces a quartic norm.  Up to an overall normalization
the quartic norm can be written as  \cite{fts}
\eq
I_4 (q) = \{\alpha \beta -T(x,y)\}^2 +6 \{ \alpha I_3(y)+\beta I_3(x) 
- T(x^{\#},y^{\#}) \}
\en
where $T(x,y) \equiv Trace (x \circ y)$ and $\#$ stands for the quadratic
adjoint map of \eja \/ which has the property \cite{kmc}
\eq
x^{\# \#}
 =I_3(x) x
\en 
The above quartic form $I_4(q)$ is invariant under the linear 
action of \esev \/ on the exceptional Freudenthal triple system. The above
realization of $ 56$ of $E_{7(7)}$ corresponds to the decomposition
\eq
56 = 27^1 + \bar{27}^{-1} + 1^3 + \bar{1}^{-3}
\en
with respect to the $E_{6(6)} \times O(1,1)$ subgroup. We should also note
that  $56$ can also be decomposed similarly with respect to the
$E_{6(2)}\times U(1)$ subgroup of \esev \/. In this case the two singlets are
complex conjugates of each other carrying opposite charges with respect to
$U(1)$. $E_{6(2)}$ has the maximal compact subgroup $SU(6)\times SU(2)$ and
 corresponds to the isometry group of the quaternionic manifold of a maximal
$N=2$ matter-Einstein supergravity truncation of the $N=8$ supergravity
in $d=4$ \cite{gstpl,adf}.
In contrast to the five dimensional case and in analogy with the Minkowskian
case we have two different classes of generic orbits with non-vanishing
quartic form $I_4$. They correspond to
\eq
\frac{E_{7(7)}}{E_{6(6)}}
\en
and to
\eq
\frac{E_{7(7)}}{E_{6(2)}}
\en

As in \cite{fm} we choose the overall sign of the quartic invariant such
that it corresponds to entropy of the BPS black holes. With this choice
the orbit corresponding to $ \frac{E_{7(7)}}{E_{6(6)}}$ has $I_4 < 0$
and the orbit corresponding to $\frac{E_{7(7)}}{E_{6(2)}}$ has $I_4 >0$.
This can be seen from the decomposition of $56$ of \esev \/ with respect
to $SU(6)\times SU(2)$ 
\eq
56= (15,1) + (\bar{15},1) + (6,2) +(\bar{6},\bar{2})+ 1 + \bar{1}
\en
and retaining the singlets \cite{adf}.

We now consider ``light-like orbits'' for which $I_4=0$. 
There are 3 distinct cases depending on the number of 
vanishing eigenvalues that lead to vanishing $I_4$. We define the
generic light-like orbit to be one for which a single eigenvalue
vanishes. The orbit in this case is given by
\eq
\frac{E_{7(7)}}{F_{4(4)}\odot T_{26}} 
\en
where $T_{26}$ is a 26 dimensional Abelian subgroup of \esev \/.
The critical light-like orbit has two vanishing eigenvalues and 
correspond to the 45 dimensional orbit
\eq
\frac{E_{7(7)}}{O(6,5)\odot ( T_{32} \oplus T_1 )}
\en
The  88 dimensional triangular subgroup of \esev \/ above
is a subgroup of the better known triangular subgroup 
\eq
O(6,6)\odot (T_{32}\oplus T_1)
\en
of \esev \/ \cite{ik,gh}. 
The doubly critical light-like orbit with  three vanishing eigenvalues is
given by the 28 dimensional quotient space
\eq
\frac{E_{7(7)}}{E_{6(6)}\odot T_{27}}
\en
 
We should note that the determination of the little groups that appear 
in the denominators of the above quotient spaces follows directly from
the various symmetry groups of \eja \/ and their different gradings
 \cite{ik,bg79,gh}.
In the next section we shall obtain the counting of the dimensions of
the orbits via a complementary procedure that follows from the normal
form for the central charge matrix and which relates orbits to BPS states
preserving different number of supersymmetries.

\section{BPS States and Supersymmetry}
\setcounter{equation}{0}
Extremal BPS black holes of $N=8$ supergravity correspond to massive 
representations of the $N=8$ supersymmetry algebra that saturate the 
BPS bound. They fall into three categories depending on whether the black hole 
background preserves $1/2,  1/4$ or $1/8$ of the original supersymmetry
\cite{fsz}.  BPS states preserving
$1/8$ supersymmetry are the only ones with non-vanishing entropy and regular
horizon.  BPS states with $1/4$ and $1/2$ supersymmetry have vanishing
 entropy \cite{kk}.
In this section we will relate the orbits of the fundamental representations
of \esix \/ and \esev \/ to these different cases. To this end we will relate
our classification to the analysis of \cite{fm}. The degeneracy of the
eigenvalues of the central charge matrix was there related to U-duality
 invariant constraints on the central charge matrix. This analysis heavily
depends on the so-called normal frame of the charge matrix which is
generically obtained by making a rotation under the automorphism group
of the supersymmetry algebra. The automorphism group of the supersymmetry
algebra essentially coincides with the maximal compact subgroup of
the duality group.        

Let us first study  the case of $d=5$. The $27$ dimensional representation
of \esix \/ corresponds to the symplectic traceless anti-symmetric tensor representation of $USp(8)$. It can be brought to a skew diagonal form via an $USp(8)$
transformation. In terms of the eigenvalues $e_i$ of this matrix the cubic
invariant takes the form   \cite{fm}
\eq
I_3=(e_1+e_2)(e_1+e_3)(e_2+e_3)
\en
We then see that the three different orbits described in the preceding section
correspond to the following three cases \cite{fm} \\
a) $I_3 \ne 0$ \\
b) $I_3=0, \qquad \frac{\partial I_3}{\partial e_i} \ne 0$  \\
c) $\frac{\partial I_3}{\partial e_i} =0 $ \\
They correspond to the cases of $1/8,1/4$ and $1/2$ supersymmetry
since in case  a) all eigenvalues are different from zero ; in case
b) two eigenvalues coincide and in case c) all three eigenvalues coincide.
Let us now count the parameters of these 3 different orbits.
We first note that the subgroup of $F_{4(4)}$ that preserves the normal 
form is $O(4,4)$ with maximal compact subgroup $SU(2)^4$ . Thus the generic 
case of $I_3 \ne 0$ involves 3 eigenvalues \cite{cy} plus 24 angles
\cite{ch} corresponding to 
\eq
\frac{USp(8)}{SU(2)^4}
\en

In case b) the little group is  $O(5,4)$ . This is a
subgroup of the triangular subgroup $O(5,4)\odot T_{16}$ of \esix \/ and again we have $2+24=26$ parameters.
In case c) corresponding to the critical orbit the little group
of the normal form is $O(5,5)$ with maximal compact subgroup 
$O(5)\times O(5) = USp(4) \times USp(4)$. Thus the number of parameters
 is one eigenvalue plus 16 angles of $\frac{USp(8)}{USp(4)\times USp(4)}$.
Note that $O(5,5)$ is a subgroup of the triangular little group of the
17 dimensional orbit.

In $d=4$ the 56 dimensional representation of \esev \/ can also be
 described by 
a complex $8\times 8$ matrix and its complex conjugate. This $8\times 8$ 
matrix can be brought to a skew diagonal form by an $SU(8)$ rotation. 
The skew diagonal form has 5 parameters \cite{cy}, an overall phase
and four real positive skew diagonal eigenvalues. The skew diagonal form 
is invariant under $O(4,4)$ which is the subgroup of $E_{6(2)}$ that preserves 
the normal form. The generic orbit then can be parametrized by five ``normal''
coordinates plus 51 angles \cite{ch} in $\frac{SU(8)}{USp(2)^4}$. In four dimensions the
extra condition for a generic state to be $1/8$ BPS with non-vanishing entropy
 is that $I_4 >0$ \cite{fm}. This is a consequence of the fact that at the
 horizon all central charge eigenvalues but the BPS mass vanish \cite{fk}. Thus
one has \cite{fk,adf}
\eq
I_4=I_{4Horizon}= M_{BPS(Horizon)}^4
\en
This selects the ``time-like'' orbit $\frac{E_{7(7)}}{E_{6(2)}}$.

Let us now consider the 3 light-like orbits.
The generic (55 dimensional) light-like orbit has four different eigenvalues
of the $8\times 8$ matrix and still preserves $1/8$ supersymmetry. The 
critical light-like orbit corresponding to $1/4$ supersymmetry has eigenvalues
 that coincide in pairs and zero overall phase \cite{fm}. The simple part of
the little group
in  this case is  $O(5,5)$ and the number of parameters
is given by the two normal parameters plus the 43 angles of
 $\frac{SU(8)}{USp(4)^2}$. The double critical orbit corresponds to four
 coinciding eigenvalues in the normal form , zero phase and $1/2$
 supersymmetry. The little group preserving this form is \esix \/ with maximal
compact subgroup $USp(8)$. The total number of parameters of the double
 critical orbit is  one normal parameter and 27 angles of 
$\frac{SU(8)}{USp(8)}$ which agrees with the results of the previous section.

 \section{The Orbits of  BPS States of $N=2$  Maxwell-Einstein Supergravity
Theories in Five and Four Dimensions}
In this section we shall extend our results to the classification of the orbits
of $N=2$ Maxwell-Einstein supergravity theories (MESGT) in five and four space-time 
dimensions.  The $N=2$ MESGT's in five dimensions were constructed by G\"{u}naydin
, Sierra and Townsend (GST) sometime ago \cite{gstpl,gstnp,gstcqg}. In the work of GST 
a complete characterization of the geometry of  five dimensional $N=2$ MESGT's theories
 as well as the geometry of four dimensional theories that are obtained by dimensional 
reduction from five dimensions was given. The geometry of the
corresponding four dimensional geometries came to be called a very special geometry 
 to distinguish them from the special geometry of general $N=2$ theories in four
dimensions \cite{specialgeom}.  Most of these coset spaces appear as moduli
spaces of untwisted sectors of orbifold compactification of type II string
theories , preserving $N=2$ supersymmetry. For a related discussion we refer
to reference \cite{cfg}.
The $N=2$ MESGT's in five dimensions and their geometries are characterized by
a cubic norm form defined by the completely symmetric tensor  $C_{IJK}$ 
that appears in front of the $FFA$ type coupling term of the vector fields.
When this cubic form is taken to be the norm form of  a Jordan algebra of
degree three, the scalar manifold of the corresponding $N=2$ theory is
 a symmetric space of the form $G/H$ where $G$ and $H$ are the reduced structure
and automorphism groups of the underlying Jordan algebra, respectively \cite{gstpl,gstnp}.
There is an infinite family of  $N=2$ MESGT's defined by reducible Jordan algebras of 
degree three whose scalar manifolds ${\cal M}$ are
\eq
{\cal M} = \frac{SO(n-1,1)}{SO(n-1)} \times SO(1,1)
\en
where $n \geq 1$ is the number of   vector multiplets coupled to
$N=2$ supergravity.  Furthermore, there exist four $N=2$  MESGT's 
defined by four simple Jordan algebras of degree 3 with irreducible norm forms.
 They are the Jordan algebras of hermitian $3\times 3$ matrices with over
the four division algebras , namely the reals ${\bf R}$, complex numbers ${\bf C}$,
quaternions ${\bf H}$ and the octonions ${\bf O}$. Their scalar manifolds
are 
\eq
\frac{SL(3,{\bf R})}{SO(3)} ; \; \frac{SL(3,{\bf C})}{SU(3)} ; \;
\frac{SU^*(6)}{USp(6)} ; \;  \frac{E_{6(-26)}}{F_4} 
\en
These theories are referred to as "magical" MESGT's since their symmetry
groups in d=5,4,3 dimensions are the groups of the famous Magic Square
of Freudenthal, Rozenfeld and Tits associated with some remarkable geometries
\cite{frt}. Except for the exceptional  MESGT defined by the exceptional Jordan algebra
$J_3^{{\bf O}}$ these theories can all be obtained by a consistent truncation
of the $N=8$ supergravity in the respective dimensions \cite{gstpl}. 
In addition to the above there exist one other infinite family
of  $N=2$ MESGT's whose scalar manifolds are symmetric spaces 
of the form \cite{gstcqg}
\eq
\frac{SO(n,1)}{SO(n)}
\en
For this family the cubic  forms are, in general, not given by the norm
forms of   Jordan algebras of degree 3 \cite{gstcqg}. 

The entropy of the BPS black holes in $N=2$ MESGT's  studied so far in the
literature is given by
 the square root of  the modulus of the cubic invariant $I_3$ determined by the tensor $C_{IJK}$
and this  is believed to hold in general \cite{fk,d5entropy}. 
One can give a complete classification
of the orbits of such states in the $N=2$ MESGT's listed above. {\it For BPS states
with $I_3$ such that all three eigenvalues are positive the orbit coincides
with the scalar manifold $G/H$ of the MESGT.}  For BPS states with $I_3$
such that two of the eigenvalues are negative and one positive the orbits are
of the form $G/ {\tilde H}$ where ${\tilde H}$ is a different real form of $H$. 
We list the possible orbits in Table I.\footnote{We should note that the corresponding 
two cases in the $N=8$ supergravity lead to isomorphic orbits. Whether both of
the distinct orbits in $N=2$ MESGT correspond to orbits of BPS states remains
to be established.}  

\begin{table}
\begin{tabular}{|c|c|}  \hline
$G/H$ &$ G/{\tilde H}$ \\ \hline
$\frac{SO(n-1,1)}{SO(n-1)}\times SO(1,1)$ &
$\frac{SO(n-1,1)}{SO(n-2,1)} \times SO(1,1)$ \\
$\frac{SL(3,{\bf R})}{SO(3)}$ &$ \frac{SL(3,{\bf R})}{SL(2,{\bf R})}$ \\
$\frac{SL(3,{\bf C})}{SU(3)}$ & $\frac{SL(3,{\bf C})}{SU(2,1)}$ \\
$\frac{SU^*(6)}{USp(6)}$ &$ \frac{SU^*(6)}{Sp(4,2)}$ \\
$\frac{E_{6(-26)}}{F_4}$ &$ \frac{E_{6(-26)}}{F_{4(-20)}} $\\
$\frac{SO(n,1)}{SO(n)}$ &$ \frac{SO(n,1)}{SO(n-1,1)} $ \\ \hline
\end{tabular}
\caption{ Above we list the possible orbits of the $N=2$
MESGT's in $d=5$ whose scalar manifolds are symmetric spaces
of the form $G/H$ . The first column corresponds to orbits with
nonvanishing cubic norm such that all three eigenvalues are positive
and coincides with $G/H$. The second column lists orbits with
positive cubic norm such that two of the eigenvalues are
negative.} 

\end{table}
We should note that in $N=2$ MESGT's in $d=5$ the scalar manifold is
a hypersurface defined by the condition that the cubic norm be equal to a non-zero
positive constant which was chosen to be one \cite{gstnp}. Similarly, the metric
of the kinetic energy term of the vector fields was given by a tensor evaluated at this 
hypersurface.   Therefore, the points where the entropy vanishes are expected to
correspond to phase transition points in these theories and/or going to the
boundary of their moduli spaces in a singular fashion \cite{vanish}. At these points
one expects the number of massless degrees of freedom to change as well.
Here we shall give the orbits corresponding to vanishing cubic norm form for the
exceptional $N=2$ MESGT in order to highlight the differences with the $N=8$
theory. 
The light-like orbit with a single vanishing  eigenvalue such that the other two 
eigenvalues are positive is
\eq
\frac{E_{6(-26)}}{O(9)\odot T_{16}}
\en
If the two non-vanishing eigenvalues have opposite signs then the lightlike
orbit is
 \eq
\frac{E_{6(-26)}}{O(8,1) \odot T_{16}}
\en
For the $N=8$ theory the corresponding orbits turn out to be isomorphic.
The critical light-like orbit for the exceptional theory with two vanishing
eigenvalues is
\eq
\frac{E_{6(-26)}}{O(9,1)\odot T_{16}}
\en

The four dimensional $N=2$ MESGT's whose scalar manifolds are symmetric
spaces $G/H$ \cite{cvp} can all be obtained from the five dimensional theories listed
above by dimensional reduction \cite{gstnp,gstcqg}.
For the theories defined by the reducible Jordan algebras the scalar manifold
in d=4 is given as \cite{gstpl,gstnp}:
\eq
\frac{SO(n,2)}{SO(n)\times SO(2)}\times \frac{SO(2,1)}{SO(2)}
\en
The scalar manifolds of the magical supergravity theories in d=4 are \cite{gstpl,gstnp}
\eq
\frac{Sp(6,{\bf R})}{U(3)} ; \; \frac{SU(3,3)}{S(U(3)\times U(3))}
; \;  \frac{SO^*(12)}{U(6)} ; \; \frac{E_{7(-25)}}{E_6 \times U(1)}
\en
and the scalar manifolds of the four dimensional $N=2$ theories
corresponding to the family (4-6) are \cite{gstcqg}
\eq
\frac{SU(n,1)}{U(n)}
\en
 Note that the manifolds $G/H$ of scalar fields in four dimensions are 
special Kahler such that $H$ has an Abelian $U(1)$ factor $H=H_0 \times U(1)$.

We recall that in $N=2$ MESGT's in d=4 the entropy formula for arbitrary
geometry can be written in a manifestly positive form as \cite{fk}:
\eq
S= \{|Z|^2 + | D_i Z|^2\}_{fix} = - \frac{1}{2} P^T M({\cal N}) P|_{fix}
\en
where $Z$ is the $N=2$ central charge and $D_iZ$ is the K\"{a}hler condensate.
The subscript $(fix)$ means that the quantities are computed for fixed moduli 
which extremize the BPS mass \cite{fk}.
The $P=(p,q)$ is a $2n+2$ dimensional vector of quantized charges and $M({\cal N})$
is a $(2n+2)\times (2n+2)$ symplectic matrix constructed in terms of the symmetric
$(n+1)\times (n+1)$ complex "kinetic' matrix ${\cal N}$ \cite{fk}. 

For those theories whose scalar manifolds are symmetric spaces $G/H$
the entropy reduces to a  $G$ invariant in the corresponding representation $P$
of $G$. Specifically,
 the entropy of the BPS black holes
in d=4 is given by the square root of the quartic invariant $I_4$
of $G$ in the representation $P$\cite{fk}. 
For the family of $N=2$ theories with the scalar manifold $\frac{SU(n,1)}{U(n)}$
the quartic form turns out to be the square of a quadratic form and the entropy
is given by this quadratic function of  electric and magnetic charges.
Requiring that
$I_4$ be positive for BPS black holes leads to two different types of orbits.
For those $N=2$ MESGT's defined by Jordan algebras of degree three
these two types of orbits are of the form
\eq
\frac{G}{H_0} 
\en
and of the form
\eq
\frac{G}{{\tilde H}_0}
\en
where $H_0$ is the maximal compact subgoup of $G$ with $U(1)$ factor
deleted and ${\tilde H}_0$ is the noncompact real form of $H_0$ which is
isomorphic to the noncompact symmetry group of the corresponding five
dimensional theory. We list these orbits in Table II.

\begin{table}
\begin{tabular}{|c|c|}  \hline
$G/H_0$ &$ G/{\tilde H}_0$ \\ \hline
$\frac{SO(n,2) \times SO(2,1)}{SO(n)\times SO(2)}$ &
$ \frac{SO(n,2) \times SO(2,1)}{SO(n-1,1)\times SO(1,1)}$ \\ 
$\frac{Sp(6,{\bf R})}{SU(3)}$ &$\frac{Sp(6,{\bf R})}{SL(3, {\bf R})} $ \\ 
$ \frac{SU(3,3)}{SU(3)\times SU(3)}$ & $\frac{SU(3,3)}{SL(3,{\bf C})} $ \\ 
$  \frac{SO^*(12)}{SU(6)}$ & $\frac{SO^*(12)}{SU^*(6)}$   \\ 
$ \frac{E_{7(-25)}}{E_6 }$ & $ \frac{E_{7(-25)}}{E_{6(-26)} }$ \\ 
$\frac{SU(n,1)}{SU(n)} $ & $\frac{SU(n,1)}{SU(n-1,1)} $ \\  \hline
\end{tabular}
\caption{ Above we list the possible orbits of the $N=2$
MESGT's in $d=4$ whose scalar manifolds are symmetric spaces
of the form $G/H$ . The first column corresponds to orbits with
nonvanishing quartic form such that all four eigenvalues are positive.
 The second column lists orbits with
positive quartic form such that two of the eigenvalues are
negative.}

\end{table}
Again for comparison with the $N=8$ theory we will give the orbits
for vanishing quartic form of the exceptional $N=2$ theory.
We find two lightlike orbits:
\eqn
\frac{E_{7(-25)}}{F_4 \odot T_{26}} \\
\frac{E_{7(-25)}}{F_{4(-20)} \odot T_{26}}
\enn
and two critical lightlike orbits:
\eqn
\frac{E_{7(-25)}}{O(9,2) \odot [ T_{32} \oplus T_1 ]} \\
 \frac{E_{7(-25)}}{O(10,1) \odot [ T_{32} \oplus T_1 ]}  
\enn
as well as two doubly critical lightlike orbits:
\eqn
\frac{E_{7(-25)}}{E_{6(-26)} \odot T_{27}} \\
\frac{E_{7(-25)}}{E_{6} \odot T_{27}}
\enn

\section{Conclusions}
\setcounter{equation}{0}
In the first part of our paper  we have determined the orbits of the exceptional groups 
corresponding
to duality symmetries of toroidally compactified string or M theories to
4 and 5 dimensions. Our analysis is classical and for the quantum theory the
relevant U-duality groups become discrete \cite{ht}. We expect our
 results can be
 extended to the discrete cases as well. 

An intriguing aspect of our results is the appearance of larger symmetries
acting non-linearly on the generalized light-cones defined by vanishing cubic
 and quartic forms. In $d=5$ this turns out to be \esev \/ that acts via linear
 fractional transformations on \eja \/. There is a a discrete subgroup 
 of \esev \/ that acts via discrete linear fractional transformations on
\eja \/. This makes it tempting to speculate that the generalized conformal
group acts as spectrum generating symmetry of the string or M theory 
toroidally compactified to $d=5$. In four dimensions we expect the analog 
of this generalized conformal group to be $E_{8(8)}$. However, $E_{8(8)}$
does not admit 3-grading with respect to any maximal rank subalgebra and hence
does not act via linear fractional transformations on $56$ of \esev \/. 
However, it has a non-linear action on a 57 dimensional space which splits
as $56+1$ under \esev \cite{bbmg}. The physical meaning of this extra singlet
is not clear. This problem may be related to the difficulty in extending 
the results on BPS black holes to 3 dimensions. We should also note that our
results can be extended to theories with less supersymmetry such as heterotic
strings and to
 dualities in space-time dimensions greater than
 five.

In the latter part of our paper we  extended our results to the classification of
the orbits of $N=2$ MESGT's whose scalar manifolds are symmetric spaces 
in five and four dimensions. We find that different signs of charge eigenvalues
lead to different orbits in this case. We also listed the orbits for vanishing
entropy  for the exceptional $N=2$ MESGT. 

The complete list of orbits of $N=2$ MESGT's in $d=4$ and $d=5$
with vanishing entropy , their physical discussion and 
the extension of the $N=8$ results to higher dimensional theories as 
well as   further details  of the results presented here will be given elsewhere
\cite{fg}.

{\bf Acknowledgements}: One of us (S.F) would like to thank C. Savoy for
 discussions
regarding the material of section 3 and J. Maldacena for several discussions.
The other (M.G) would like to thank H. Nicolai for numerous enlightening
discussions.

\end{document}